\documentclass{elsart3p}
\usepackage{graphicx}

\usepackage{amssymb}\usepackage{graphics}

\begin{document}

\begin{frontmatter}

\makeatletter

\providecommand{\LyX}{L\kern-.1667em\lower.25em\hbox{Y}\kern-.125emX\@}





\title{Antiferromagnetic correlations near the lower edge of superconducting dome in YBCO6+x}


\author[a]{Z. Yamani},* \ead{Zahra.Yamani@nrc.gc.ca}
\author[a,b]{W.J.L. Buyers},
\author[c]{F. Wang},
\author[c]{Y-J. Kim},
\author[b,d]{R. Liang},
\author[b,d]{D. Bonn},
\author[b,d]{W.N. Hardy}

\address[a]{National Research Council, Canadian Neutron Beam Centre, Chalk River Laboratories, Chalk River ON K0J
1J0, Canada}
\address[b]{Canadian Institute of Advanced Research, Toronto, ON M5G 1Z8, Canada}
\address[c]{Department of Physics, University of Toronto, 60 St. George Street, Toronto
Ontario M5S 1A7, Canada}
\address[d]{Department of Physics and Astronomy, University Of British Colombia, 6224 Agricultural Road,
Vancouver BC V6T 1Z1, Canada}


\begin{abstract}

Neutron scattering from high-quality YBCO6.334 single crystals with a T$_c$ of 8.4 K shows that there is no
coexistence with long-range antiferromagnetic order at this very low, near-critical doping of $\sim$0.055, in
contrast to claims based on local probe techniques. We find that the neutron resonance seen in optimally doped
YBCO7 and underdoped YBCO6.5, has undergone large softening and damping. It appears that the
overdamped resonance, with a relaxation rate of 2 meV, is coupled to a zero-energy central mode that grows
with cooling and eventually saturates with no change at or below T$_c$. Although a similar qualitative
behaviour is found for YBCO6.35, our study shows that the central mode is stronger in YBCO6.334 than YBCO6.35.
The system remains subcritical with short-ranged three dimensional correlations.

\end{abstract}

\end{frontmatter}

\section{Introduction}
\label{labelOfFirstSection}

Rich phase diagram of YBa$_2$Cu$_3$O$_{6+x}$ (YBCO6+x) system is accessible over the entire range of doping,
but relatively few studies exist at low doping near the critical onset for superconductivity. One particularly
important question is whether long-ranged spin ordering coexists with superconductivity or competes with it.
Neutron scattering experiments have allowed us to unambiguously determine the length and time scales of the
magnetic structure as it evolves with temperature and doping. Since there is little consensus on the nature of
the precursor phase from which the superconducting (SC) phase emerges when holes are doped into a planar array of correlated
S=1/2 Cu spins, such results are needed to determine whether there is a novel phase between the antiferromagnetic (AF) and SC
phases.

\section{Results}
\label{labelOfSecondSection} We have studied very lightly doped single crystals of YBCO6.334 with T$_c$ of
only 8.4 K. The crystals are well-annealed and in the orthorhombic phase. Neutron scattering experiments were
performed at the C5 spectrometer at NRU reactor Chalk River with PG(002) as both monochromator and analyzer.
Elastic and inelastic properties of the sample aligned in the (HHL) plane were studied in the temperature
range 1.5 K to 300 K and up to 15 meV energy transfer.

\begin{figure}[tbh]
\begin{center}
\vskip 0cm \resizebox{.8\linewidth}{!}{\includegraphics{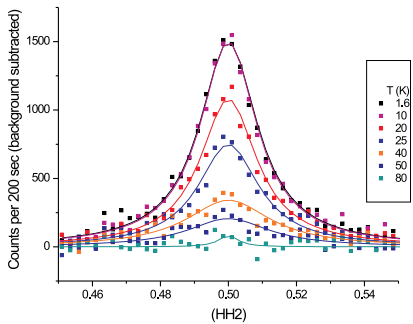}}\vskip 0cm \caption []{Temperature
dependence of the elastic scattering observed around the AF (0.5 0.5 2) position. The data is corrected for
the background by subtracting the 100 K data. The solid lines are fits to Lorentzians. The peak intensity
grows by an order of magnitude on cooling without a corresponding narrowing of the peak width. The resolution
width is 0.015 rlu.} \label{HH2}
\end{center}
\end{figure}

Elastic scans along the [HH0] and [00L] directions have been performed around the AF position (0.5 0.5 2) as a
function of temperature. We find that an elastic signal distributed about the AF position grows strongly on
cooling below 60 K. The peak widths along both directions remain broad even at 1.5 K and show that spin
correlations extend over about $\sim$20 cells in the ab-plane. There is no transition to a long-range AF
state. The elastic scans, corrected for background, along [HH0] are shown in Fig.~\ref{HH2} at different
temperatures. For the scans along [0.5 0.5 L] we find that the scattering is peaked at integer L-values but
with an extremely short correlation scale of $\sim$2 unit cells~\cite{yamani07}. We conclude that an incipient three
dimensional (3D) AF pattern coexists with superconductivity, but not AF long-ranged order.

\begin{figure}[tbh]
\begin{center}
\vskip 0cm \resizebox{.8\linewidth}{!}{\includegraphics{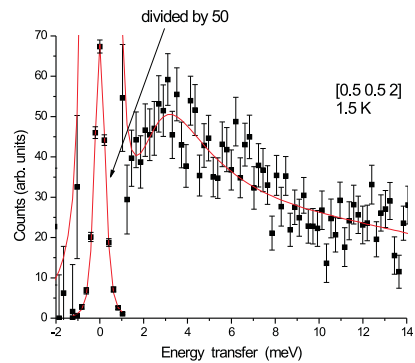}}\vskip 0cm \caption []{Inelastic
spectrum observed at 1.5 K at the AF position (0.5 0.5 2). The average of the data measured at (0.3 0.3 2) and
(0.7 0.7 2) is used as background. The solid line is a fit to a resolution-limited Lorentzian at zero energy
and to a broad damped response with a $\sim$2 meV relaxation rate. } \label{inelastic}
\end{center}
\end{figure}

The inelastic spectrum at (0.5 0.5 2) and 3 K is shown in Fig.~\ref{inelastic}. The spectrum contains two
energy scales: a very slow response characterized by a central peak, resolution limited to a width less than
0.08 meV at low temperatures, and a fast component giving rise to the broad peak centred at 2 meV. Scans in q
indicate that the correlations associated with the central peak and the inelastic broad peak extend over a
similarly short ranges. Although qualitatively similar to YBCO6.35~\cite{stock06}, we find that the spectral
weight of the central mode relative to the inelastic peak is larger by a factor of more than two in YBCO6.334
compared to YBCO6.35 as suggested also by the larger correlation range. The low energy spectra of YBCO6.35 and
YBCO6.334 are different from YBCO6.5~\cite{stock05,stock04}, where no central mode is observed, a well-defined
commensurated resonance occurs at high energies while at lower energies the inelastic fluctuations are
incommensurate. Since it is the only identifiable spectral feature, we believe that the low energy damped
commensurate response that we observe represents the spectral weight of the resonance, that has been moved to
very low energies.  It is the soft spin mode of the superconducting phase as it tracks T$_c$ as a function of
doping as suggested in Ref.~\cite{buyers06}. In addition, it appears that this behaviour differs from
La$_{2-x}$Sr$_x$CuO$_4$ for which incommensurate modulations occur with a wave vector proportional to doping,
possibly indicating a different precursor phase for superconductivity.

Short correlation lengths for the central mode indicate that the doped holes have produced extensive regions
which break up the AF coupling and likely create ferro correlations through the spins on the oxygen
neighbours.

\end{document}